\documentstyle[epsf,prl,aps,multicol]{revtex}

\begin{document}

\title{
Dephasing due to background charge fluctuations
}

\author{ 
Toshifumi Itakura and Yasuhiro Tokura}

\address{
   NTT Basic Research Laboratories, NTT Corporation \\
  3-1, Morinosato Wakamiya, Atsugi-shi,
  Kanagawa Pref., 243-0198 Japan \\
    Tel: +81-46-240-3938
    Fax: +81-46-240-4726  \\
    E-mail: itakura@will.brl.ntt.co.jp
}

\date{recieved March 18, 2002. accepted on March 20, 2003}

\maketitle

\draft
\begin{abstract}
In quantum computation,
quantum coherence must be maintained during  gate operation.
However, in physical implementations,
         various couplings with the environment 
         are unavoidable
         and can lead to a dephasing of a quantum bit (qubit).
The background charge fluctuations are
         an important dephasing process,
         especially in a charge qubit system. 
We examined the dephasing rate of a qubit due to
            random telegraph noise. 
Solving stochastic differential equations,
    we obtained the dephasing rate of a qubit
    constructed of a coupled-dot system;
    we applied
    our results to the
    charge Josephson qubit system.
We examined the dephasing rates due to      
      two types of  couplings between the coupled-dot system
      and the background charge, namely, 
      fluctuation in the tunnel coupling constant and
      fluctuation in the asymmetric bias.
For a strong coupling condition,
      the dephasing rate was inversely proportional 
      to the time constant of the telegraph noise.
When there is fluctuation in the tunnel coupling constant,
     Gaussian decay occurs in the initial regime.
We also examined the rate of dephasing due to many impurity sites.
For a weak coupling condition with fluctuation in the asymmetric bias, 
      the obtained dephasing rate coincided with
      that obtained by the perturbation method
      using  the spectral weight of a boson thermal bath,
      which is 
      proportional to the inverse of the frequency.

\end{abstract}

\draft



\begin{multicols}{2}

\section{Introduction}

Efforts to implement  quantum computation have recently intensified.
The application of the quantum bit (qubit) to solid-state materials,
such as superconducting Josephson junctions
\cite{Nakamura}
and quantum dots
\cite{Loss,Tanamoto,Miranowicz},  is particularly  promising,
       because
       these implementations have the advantage of  scalability.
In a coupled-dot system,
        for example,
        the localized states in the left  and right dots are treated
        as a basic two-level system,
        in which the tunnel coupling
        between the two dots
        constructs a quantum superposition of the dots.
This superposition
        manifests itself in coherent quantum oscillation
       (Rabi oscillation),
        and transition can be induced between the superposed states
\cite{Fujisawa_NT}. 
A nanometer-scale superconducting electrode connected to a reservoir
    via a Josephson junction constitutes an artificial two-level 
    system in
   which the charge states, coupled by tunneling, differ by $2e$,
    where $e$ is the electronic charge.
    This system has shown clear  Rabi oscillation    
\cite{Nakamura}.

Quantum coherence must be maintained during quantum gate operation.
Dephasing, characterized by the dephasing time,
        originates from various couplings
         between the qubit and the environment.
When the 
        qubit is implemented in a 
        solid-state system, the effects of phonons and 
        electromagnetic 
        and background charge fluctuations (BCFs)
        are important in the dephsing process.
The effect of phonons has been examined in semiconductor quantum dots
        as the source of the dephasing accompanying dissipation
        \cite{Fujisawa_SC};
        the effect of electromagnetic fluctuation
        in Josephson junction qubits has been extensively studied
          \cite{Schon}.  
However, BCFs have 
     not yet been examined systematically,
       in spite of their importance in the dephasing process.

BCFs 
         have been observed in many systems
\cite{Devoret,Martinis,Lyon,Zorin}.
In nanoscale systems,  they are
     the electrostatic potential fluctuations
     due to the dynamics of electrons or holes
     trapped at impurity sites.
In particular, the charge of a single impurity 
     fluctuates
     with the  Lorentzian spectrum form,
     which is 
     called
     "random telegraph noise"
     in the time domain
\cite{Lyon,Fujisawa_BC}.     
The random 
      distribution of the positions of such impurities
       and their time constants  
     lead to  BCFs or 1/f noise
\cite{BC}.
In solid-state charge qubits,
    these  BCFs
    lead to  a  dynamical electrostatic disturbance and
    hence the dephasing.
The effect of 1/f noise on a charge Josephson
     qubit has been examined theoretically -
     the interaction between the qubit and environment has been treated
     by the perturbation method
     \cite{Fazio,Shnirman},
     by Gaussian approximation \cite{Nakamura_CE}
     and by  the path integral method within a spin-Boson model
     \cite{Fazio,Shnirman}.
When fluctuating impurities exist in the substrate,
      not in the junction \cite{Zorin},
      the coupling between the qubit and BCFs
      is weak,
      and 
      the perturbation method is sufficient.
However, when the interaction between the qubit and environment is strong,
      methods that go beyond  perturbation are needed.
    
In this study, we 
       investigated 
       how the electrostatic disturbance
      of time constant $\tau_0$ coming from
      a single impurity
      affects the quantum coherence of a qubit
      irrespective of the strength of the qubit-impurity coupling.
We also examine the effect of many impurity sites.
This approach is in clear contrast with previous  ones  
        \cite{Nakamura_CE,Shnirman},
        in which the phenomenological spectral weight 
        of the boson thermal bath 
        was used to characterize the effect of BCFs.

We consider  two types of couplings between the qubit and the environment:
pure dephasing  
      and 
      dephasing accompanied with relaxation of the population
      \cite{comment}.
In symmetrical coupled-dot systems,
       the former corresponds to a fluctuation in the tunnel coupling constant,
       and the latter to that in the asymmetric bias
\cite{Galperin,Nazarov}.
 The mapping from a coupled-dot system to the Josephson charge qubit is 
        discussed in Sec. VI. 
By using the method of stochastic differential equations,
     we obtain  analytically  the dephasing rate,
        which is shown to be always smaller than $\tau_0^{-1}$.
It should be noted that this dephasing process does not mean 
        the qubit 
        becomes entangled with the environment,
        but rather
        it 
        means
        the stochastical evolution of an external classical field,
        suppressing the off-diagonal density matrix elements 
        of the qubit after being averaged over statistically distributed
         samples.

Section  II defines the Hamiltonian of the system.
Section III explains the method of stochastic differential equations.
The fluctuations in tunnel coupling
          and  asymmetric bias are examined
          in Secs. IV and V, respectively.
Section VI is devoted to discussion,
       including the effect of many impurities.
       Section VII summarizes the paper.                 


\section{Hamiltonian}

The qubit and
      the effect of a single impurity 
      are examined  in terms of the following Hamiltonian:
\begin{eqnarray}
{\cal H} &=& H_{qb}  + H_{qb-imp},  \\
  H_{qb} &=& \frac{\hbar \Delta}{2}
   ( c^{\dagger}_L c_R + c^{\dagger}_R c_L )  +
            \frac{ \hbar \epsilon}{2}
             ( c^{\dagger}_L c_L - c^{\dagger}_R c_R ),
\end{eqnarray}
where  $c^{\dagger}_{L,R}$ and $c_{L,R}$ are the 
        creation and annihilation operators
        of the left  and  the right dots, assuming a single level
        for each dot, as shown in Fig. 1(a).
The $\Delta$ is the tunnel coupling 
        between
        the dots, and
        $\epsilon$ is the asymmetric bias between them.
The interaction between a qubit and the charge at the impurity site
        is described by the following Hamiltonian: 
\begin{eqnarray}
  H_{qb-imp} &=& \frac{\hbar J_{T}}{2}
   ( c^{\dagger}_L c_R + c^{\dagger}_R c_L ) 
  2 ( d^{\dagger} d - 1/2 ) \nonumber \\
  &+&
  \frac{\hbar J_{B}}{2} ( c^{\dagger}_L c_L - c^{\dagger}_R  c_R )
  2 ( d^{\dagger} d - 1/2 ),
\end{eqnarray}
       where $J_T$ is the magnitude of the fluctuation in the
        tunnel coupling,
         $J_B$ is the magnitude of the fluctuation in the asymmetrical bias,
 and $d^{\dagger}$ and $d$ are the creation and annihilation operators
        of the charge at the impurity site, respectively.

\begin{figure}
\hspace{1truecm}
\vspace{-0.5truecm}
\center
\centerline{\epsfysize=3in
\epsfbox{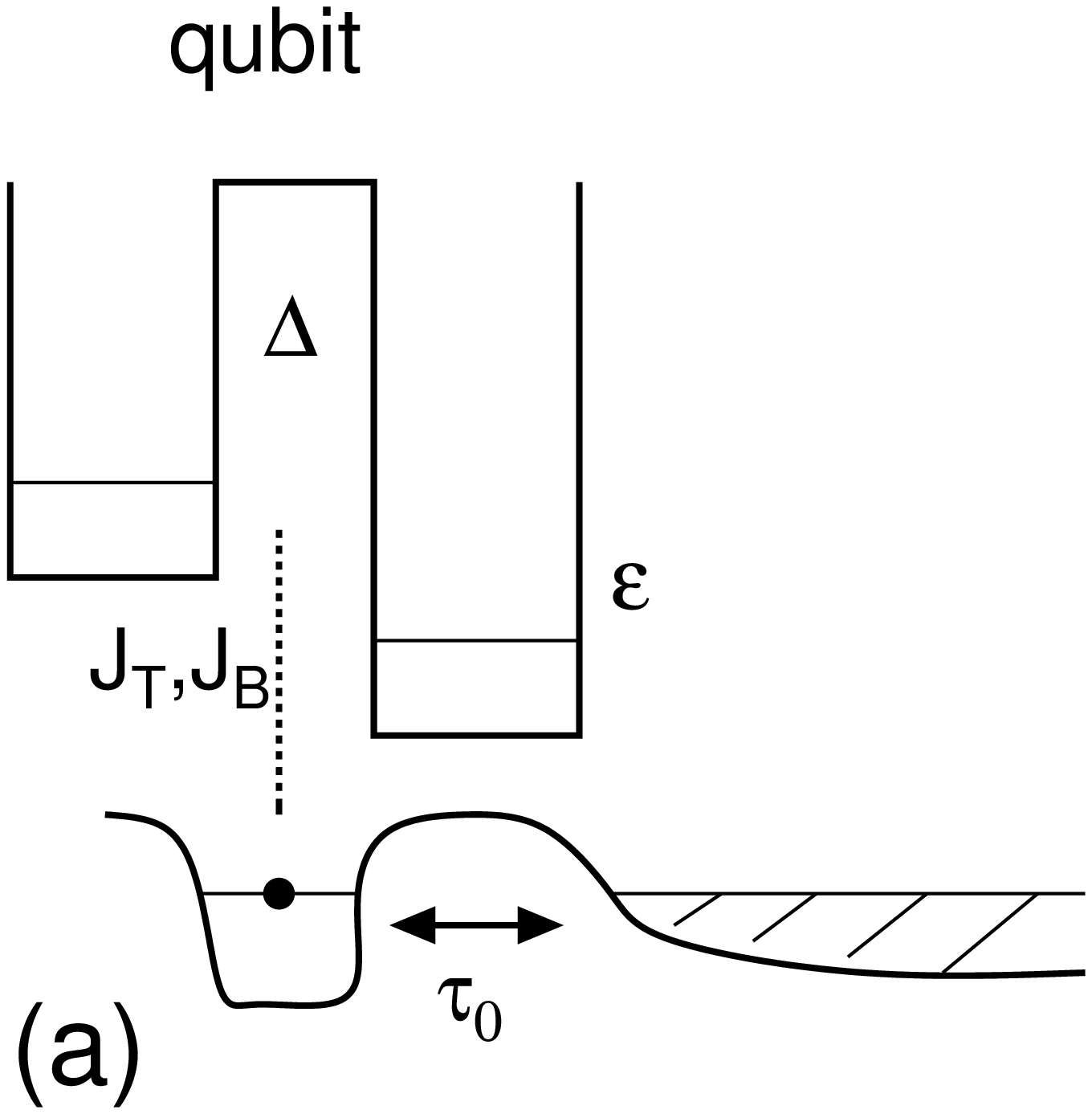}
}
\centerline{\epsfysize=2in
\epsfbox{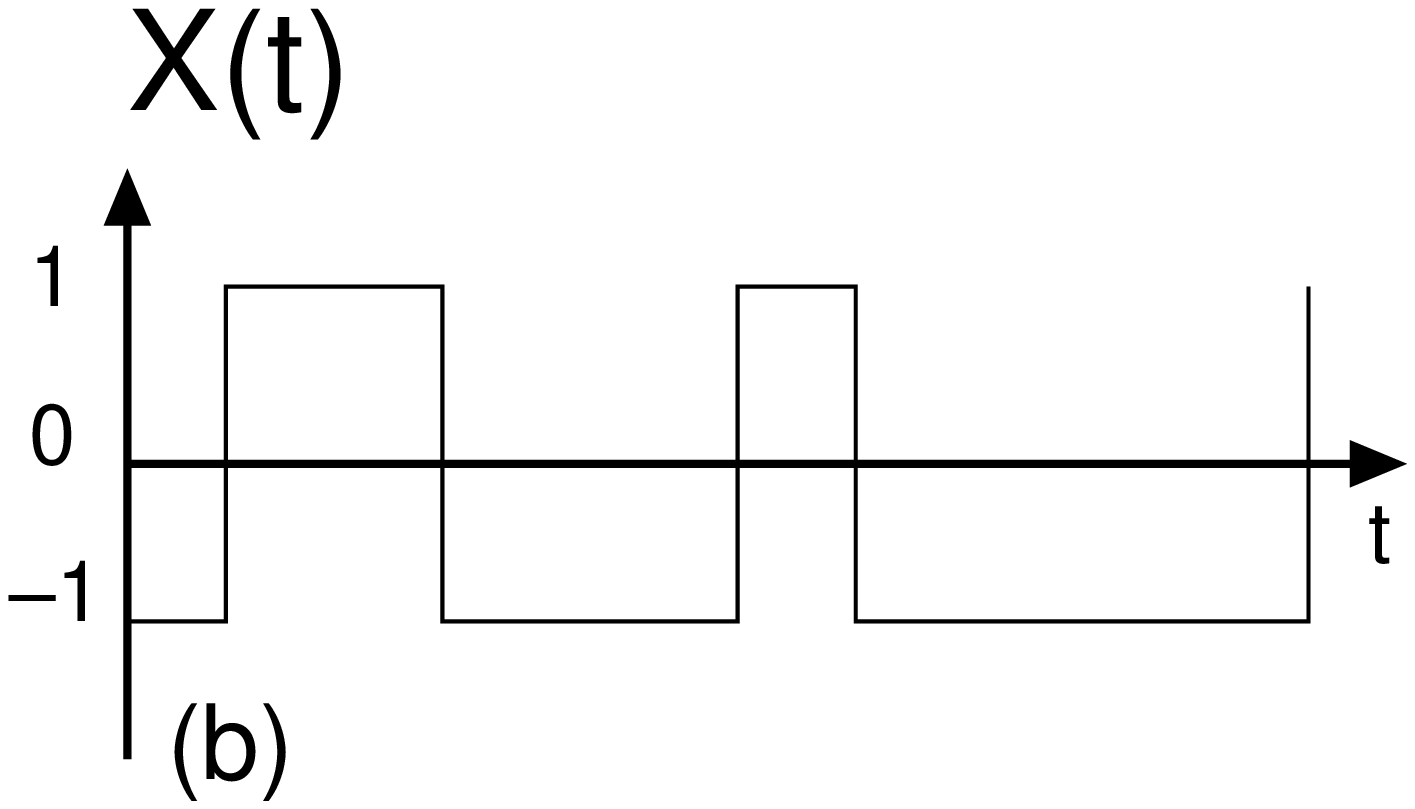}
}
\vspace{-0.5truecm}
\begin{minipage}[t]{8.1cm}
\caption{
(a) Schematic diagram  of  coupled dots and  environment
constituting an impurity site and the electron reservoir.
(b)
Example time sequence of
 random telegraph noise.
}
\label{figure1}
\end{minipage}
\end{figure}

We assume that the time evolution of  statistical variable
       $ X (t)  ( = 2 < d^{\dagger} d - 1/2 >)$
            is a Poisson process.
Assuming a  strong coupling between the charge at the impurity site
       and 
       the nearby electron reservoir,
       the dynamics of the charge  induces 
         not  Gaussian white noise,            
        but  random
        telegraph noise
\cite{Fujisawa_BC,Galperin,Nazarov}.    
We therefore consider the effect of random telegraph noise with 
              characteristic time constant  $\tau_0$,
            where the statistical variable 
           $X(t)$ takes the value  1 or -1 (Fig. 1(b)).

The time constant is determined by
          the barrier height of the electron or hole trap and the temperature,
        like
        $\tau_0 = A e^{ W / k_B T}$, where $W$, $k_B$, $T$, and $A$
         are the activation
        energy of the impurity potential, Boltzmann's constant,
        the temperature,
        and the temperature-independent prefactor, respectively
        \cite{Imry}.
When the temperature decreases, $\tau_0$ 
             becomes longer. 
The telegraph noise 
        has been experimentally observed to take a value
        of 1 or -1
        with asymmetric probabilities,
        which
        arises from the difference between the Fermi energy
        of the electron reservoir and the energy level of the 
        impurity sites.
        \cite{Fujisawa_BC}.
To include this asymmetric weight,
        we introduce asymmetric probabilities $p_u$ and $p_d$,
        which relate the asymmetric transition rates 
        for the process from -1 to 1  
        ( $\tau_u^{-1}  = ( p_u \tau_0 )^{-1}$ ) 
        to those of the opposite process
        ( $\tau_d^{-1} = ( p_d \tau_0 )^{-1}$).

We neglect the backaction from the qubit to the 
              charge
              at the impurity site,              
    so this environment reduces to a 
     classical stochastic external field.
We also assume that the temperature is high enough
         for
         the effect of the quantum fluctuation of the charge
         between the impurity site and
         electron reservoir to be neglected.

For compactness, we rewrite the Hamiltonian 
        in terms of the Pauli matrices while rotating basis $\pi/2$
        from the basis of localized states in the left and right dots
         to bonding-antibonding basis:
\begin{equation}
            \label{eqn:HM}
{\cal H} =
\frac{ \hbar \Delta}{2} \sigma_z + \frac{ \hbar \epsilon}{2} \sigma_x
         + \frac{ \hbar J_T }{2} \sigma_z X (t)
                                 + \frac{ \hbar J_B }{2} \sigma_x X (t).
\end{equation}
In the following, we consider only the case of 
     a symmetrical  coupled-dot  system, $\epsilon = 0$,
     in which the effect of the bias fluctuation due to the dephasing
     starts only from the second order, $J_B^2$, as shown
     in the following, and is less effective in the perturbation regime.
As reduced Hamiltonian Eq. (\ref{eqn:HM}) suggests,
 the present results can also be applied to other quantum two-level systems
       in which  telegraph-type fluctuation exists
(see Sec. VI).

\section{Method}

We are interested in the time-evolution of the 
  qubit's two-by-two density matrix,
  $\rho (t)$, with an arbitrary initial condition at $t=0$, $\rho (0)$.
If BCF is absent,
   starting with the initial condition that the left dot is occupied,
        for example,
        the density matrix at $t=0$ is given by
\begin{eqnarray}
     \rho (t=0) = W (\frac{\pi}{2}) 
      \left(
      \begin{array}{cc}
    1
      &
    0 \\
    0 &
    0 \end{array}
    \right) 
    W^{\dagger} (\frac{\pi}{2}),
\end{eqnarray}
where $W( \frac{\pi}{2} )$ is the matrix of rotation 
from the localized basis to
 the bonding-antibonding basis.
At time t,
\begin{eqnarray}
     \rho (t) &=& 
        e^{-i \frac{t}{2}  \Delta \sigma_z  } W (\frac{\pi}{2})
          \left(
      \begin{array}{cc}
    1
      &
    0 \\
    0 &
    0 \end{array}
    \right)  W^{\dagger} ( \frac{\pi}{2} )
    e^{ i \frac{t}{2}  \Delta \sigma_z }
     \\
 &=&  W ( \frac{\pi}{2} )
      \left(
      \begin{array}{cc}
      \frac{1 + \cos \Delta t}{2}
      &
      i \sin{ \Delta t} \\
    - i \sin{ \Delta t} &
    \frac{1 - \cos \Delta t}{2}
    \end{array}
    \right) 
    W^{\dagger} ( \frac{\pi}{2} ),
\end{eqnarray}
where the bases of the inner matrix 
     are the left and right occupancy states.    
Therefore, the density matrix shows Rabi oscillation with  
       frequency  $\Delta$.    

In the following,   
     we 
     keep the matrix indices in the bonding-antibonding basis.
To examine the instantaneous potential change,
we use the method of stochastic  differential equations
\cite{Burshtein}.
The density matrix averaged over
    all possible
     sequences of telegraph noise
    can be represented as a series:
\begin{eqnarray}
       \label{eqn:emeqn}
 \rho (t) e^{t/\tau_0} &=& \sum_{k=0}^{\infty}
                    \frac{1}{\tau_0^k}
                    \int_0^t d t_k
                    \int_0^{t_k} d t_{k-1}
                    \dots \int_0^{t_2} d t_1 \nonumber \\
               &\times& \int_X d W ( X_k )
                \int_X d W ( X_{k-1} )
                \dots \nonumber \\
                &&
                \int_X d W ( X_0 )
                \rho (t,t_k),
\end{eqnarray}
     where 
     $dW(X)$ is the distribution of $X$
     with the probability of $p_u$ for $X=1$ and
     that of $p_d$ for $X=-1$,
    with the constraint
    $p_u + p_d =1$.
The density matrix before the ensemble average,
    $\rho (t,t_k)$, is given by
\begin{eqnarray}
   \rho ( t, t_k ) &=&
   S ( X_k ; t, t_k ) \cdots S ( X_1 ; t_2, t_1 ) 
   S ( X_0 ; t_1, 0) \nonumber \\
   \times & \rho (0) & S^{-1} ( X_0 ;t_1,0)
   S^{-1} ( X_1 ; t_2, t_1 )
   \cdots S^{-1} ( X_k ; t, t_k),
\end{eqnarray}
  where       $S(X ; t, t')=S(X,t-t')$ is the unitary 
         time evolution operator,
          which 
         is determined by 
\begin{equation}
 i \hbar \frac{d S (X, t) }{dt} = {\cal H} (X) S (X,t).
\end{equation}
The explicit form of 
   $S(X,t)$ is given 
   by
\begin{eqnarray}
         \label{eqn:smatrix}
&& S (X,t)    \nonumber \\
     &=& \left(
      \begin{array}{cc}
    \cos \frac{1}{2} a t 
    -i ( \frac{\Delta + J_T X}{a})
    \sin \frac{1}{2} a t
      
     &
    - i \frac{J_B X}{a} \sin \frac{1}{2} a t \\
    - i \frac{J_B X}{a} \sin \frac{1}{2} a t &
    \cos \frac{1}{2} a t 
    + i \frac{\Delta + J_T X}{a}
    \sin \frac{1}{2} 
     a t 
    \end{array}
    \right) , \nonumber \\
\end{eqnarray}
where $a=\sqrt{(\Delta + J_T X)^2 + J_B^2}$.
Equation (\ref{eqn:emeqn})
       can be rewritten in terms of the integral equation
\begin{eqnarray}
       \label{eqn:prorho}
\rho (t) e^{\tau/\tau_0} &=& \int_{X} 
       S( X; \tau, 0) \rho (0) S^{-1} ( X ; \tau, 0)
       d W ( X ) \nonumber \\
       + & \frac{1}{\tau_0} & \int^{\tau} e^{t/\tau_0} \int_{X}
       S ( X; \tau, t) \rho (t) S^{-1} ( X ; \tau, t)
       d W ( X ) dt. \nonumber \\
\end{eqnarray}
Using $S(X; \tau, t)$, we define  matrix $R^{im} (\tau,t)$
          as follows:
\begin{eqnarray}
               \label{eqn:rmatrix}
R_{lk}^{im} (\tau,t) &=& \int_{X} S_{ik} (X ; \tau, t)
     S_{lm}^{-1} (X ; \tau, t) d W ( X ),
   \\
   R_{lk}^{im} (\tau,t) &=& ( R_{kl}^{mi} )^* (\tau ,t ).
\end{eqnarray}
We can then reduce Eq. (\ref{eqn:prorho})
       to the following compact form:
\begin{eqnarray}
   \rho_{im} ( \tau ) &=& e^{- \tau/\tau_0} 
   {\rm Tr} [ R^{im} ( \tau, 0) \rho( 0)]
   \nonumber \\
   + & \frac{1}{\tau_0} \int_0^{\tau} &
   dt {\rm exp} [ - \frac{\tau-t}{\tau_0} ]{\rm Tr}
   [R^{im} ( \tau, t ) \rho (t)].
\end{eqnarray}

\section{Fluctuation in  tunnel coupling}
  
First, we consider the case of fluctuation in  tunnel coupling
($J_T \ne  0, J_B=0$).
Since the interaction Hamiltonian  commutes with 
   $H_{qb}$, the  environment leads to
   pure dephasing without energy dissipation.
We derive $S(X,t)$ and $R_{lk}^{im} (t)$ 
           from Eqs. (\ref{eqn:smatrix}) and (\ref{eqn:rmatrix}) 
           as follows:
\begin{eqnarray} 
  S_{kl} ( X, \tau - t ) &=& {\rm exp} [  \frac{i}{2} (  \Delta ( \tau - t )
             -  
             J_{T} X (\tau -t) ) (-1)^k ] \delta_{kl},
             \nonumber \\
             \\
  R_{lk}^{nm} ( \tau-t )&=& \int_{X} {\rm exp} [ i  \Delta ( \tau - t )
           +  i J_T X (\tau - t) ]
           \nonumber \\
           & \times & \delta_{nk} \delta_{lm} d W ( X ). \nonumber \\
\end{eqnarray}      
As a result, the off-diagonal element of the density matrix
       obeys the following integral equation: 
\begin{equation}
              \label{eqn:RINT}
  \sigma ( \tau ) = e^{-\tau/\tau_0} R ( \tau )
    + \frac{1}{\tau_0}
    \int_0^{\tau}
    R( \tau - t) {\rm exp} [ - \frac{\tau-t}{\tau_0} ] \sigma (t)dt,
\end{equation}
 where    $\sigma (\tau ) = e^{-i \Delta \tau}
          \rho_{12} ( \tau ) / | \rho_{12} (0) |$
          is a normalized off-diagonal element of the qubit density matrix
          measured 
          in a 
          Rabi oscillation frame,
 and 
\begin{equation}
R( \tau - t ) =  p_u {\rm exp} [ i  J_T (\tau -t )]
               + p_d {\rm exp} [ - i J_T (\tau-t )].
\end{equation}
Equation (\ref{eqn:RINT}) can be rewritten as              
   a differential equation:
\begin{equation}
   \label{eqn:ODEPURE}
  \sigma'' ( \tau ) + \frac{1}{\tau_0} \sigma' ( \tau )
  + (  J_T^2 +  i ( \frac{ p_u - p_d }{\tau_0} ) J_T ) \sigma ( \tau )
   = 0.
\end{equation}
The initial conditions are 
 $\sigma (0) = e^{i \phi}=\rho_{12} (0) /|\rho_{12} (0)| $
  and $\sigma'(0) =0$.      
As a result of coupling with the environment,
       the off-diagonal element of the density matrix
       decays as a function of time.
When the real part of the two roots 
      of the characteristic equation of Eq. (\ref{eqn:ODEPURE}) 
      almost completely degenerates,
      the short-time 
      behavior for $t < \min ( \frac{\sqrt{2}}{J_T} , 3 \tau_0 )$
      (initial regime)
      is not a simple exponential decay. 
In this initial regime, the off-diagonal element of the density matrix
      shows Gaussian decay,
      $\sigma (t) \sim \sigma (0) ( 1 - J_T^2 t^2 / 2 + \cdots) \simeq 
       \sigma (0) e^{-J_T^2 t^2 /2 }$, 
      irrespective of the asymmetric probabilities, $p_u$ and $p_d$.
The decay of the off-diagonal element of the density matrix becomes
 exponential for the asymptotic regime,
       $ t \gg  (
         1  /( \sqrt{ | \frac{1}{\tau_0^2} - 4 J_T^2 }| ),$
          when $p_u=p_d$. 
For $J_T \tau_0 \ll 1$,
 this criterion is obtained when
  one of the two exponential decay terms becomes negligibly small.
For $J_T \tau_0 \gg 1$, the time constants of the envelope
         of the two dumped oscillating terms are the same,  $1/ (2\tau_0)$.
Exponential decay
         appears after  the inverse 
         of the oscillation frequency: 
         $ 1/ \sqrt{|4J_T^2 - \frac{1}{\tau_0^2}|}$.         
When $p_u=p_d$ and $J_T \tau_0 = \frac{1}{2}$, one obtains,
        $\sigma (t) = e^{-\frac{t}{4 \tau_0}}(1+\frac{t}{2 \tau_0})$,
        where
        the dephasing can never be a simple exponential decay. 

The time constant of this exponential decay 
       corresponds to the dephasing time, $T_2$.
For   $p_u \ne p_d$
\cite{Galperin},
\begin{equation}
            \label{eqn:T2PURE}
T_{2}^{-1} = \frac{1}{2} \Re  [ \frac{1}{\tau_0}
           - \sqrt{  ( \frac{1}{\tau_0} )^2
           - 4 J_{T}^2 
           - 4 i   ( \frac{p_u - p_d}{\tau_0} ) J_T  } ].
\end{equation}
  
\begin{figure}
\hspace{1truecm}
\vspace{-0.5truecm}
\center
\centerline{\epsfysize=3in
\epsfbox{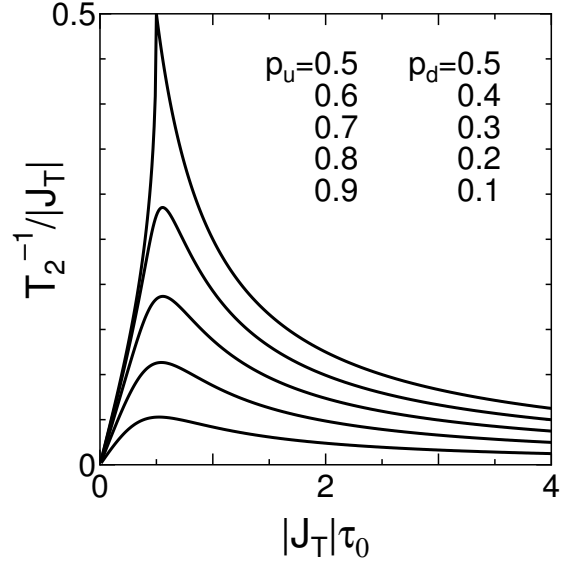}
}
\vspace{-0.5truecm}
\begin{minipage}[t]{8.1cm}
\caption{
 Dependence of dephasing rate $T_{2}^{-1} / | J_T |$
 on tunnel coupling constant $|J_T|\tau_0$
 for various values of $p_u$ and $p_d$
 with $J_T \ne 0$ and $J_B=0$.
}
\label{figure1}
\end{minipage}
\end{figure}
\noindent
Figure 2 shows the $|J_T| \tau_0$ dependence of dephasing rate $T_2^{-1}$
     for $p_u = 0.5$, 0.6, 0.7, 0.8, and 0.9.
In the two limits of weak and strong coupling, we have
\begin{eqnarray}
  T_2^{-1}= \left\{
         \begin{array}{rl}
         ( 1 - (p_u - p_d)^2 ) J_T^2 \tau_0, 
         & \quad \mbox{$1/\tau_0 \gg |J_T|$}. \\
         ( 1 - |p_u - p_d|)/ 2 \tau_0,
         & \quad \mbox{$1/\tau_0 \ll |J_T|.$}
         \end{array}
         \right.
\end{eqnarray}
Namely, for a fixed $|J_T|$, when  $\tau_0$ increases from 0,
    $T_2^{-1}$ first increases and then decreases.
It has a single maximum of $(1-\sqrt{|p_u-p_d|}) | J_T |$ 
      when $\tau_0=1/(2|J_T|)$; therefore, for any parameters,
      $T_2 > 2 \tau_0$. 
Changing  weight $p_u$ to make it 
       more asymmetric reduces $T_2^{-1}$.
In the limit of $p_u \to 0$ or 1, the environment is nearly frozen,
      so the dephasing time becomes infinity.

\section{Fluctuation in asymmetric bias}
           
Next we examine the effect of bias fluctuation ($J_T = 0,J_B \ne 0$).
We 
      consider only the  case of symmetrically weighted
      telegraph noise ($p_u=p_d=1/2$) for simplicity.
In this model, $H_{qb}$ and 
 the interaction Hamiltonian
 do not commute
 and the dephasing process is accompanied by a relaxation of the population.
The unitary operator is thus
\begin{eqnarray}
 && S (X,t) \nonumber \\ 
 &=& 
    \left(
    \begin{array}{cc}
   \cos \frac{1}{2} \Omega t 
    -i \frac{\Delta}{\Omega}
    \sin \frac{1}{2} \Omega t 
     
     &
    - i \frac{J_B X}{\Omega} \sin \frac{1}{2} \Omega t\\
    - i \frac{J_B X}{\Omega} \sin \frac{1}{2} \Omega t &
    \cos \frac{1}{2} \Omega t 
    + i \frac{\Delta}{\Omega}
    \sin \frac{1}{2} \Omega t
     
     \end{array} 
     \right),
\end{eqnarray}
where $\Omega=\sqrt{\Delta^2 + J_B^2}$ is the nutation frequency.
After averaging,
       the few 
       non-vanishing elements of  matrix 
       $R_{lk}^{im} ( t )$
       have the form
\begin{eqnarray}
  R_{11}^{11} ( \tau - t) &=& R_{22}^{22} ( \tau -t )= 1 - P_{12} ( \tau -t), 
  \nonumber \\
  R_{22}^{11} ( \tau - t ) &=& R_{11}^{22} ( \tau -t ) = R_{12}^{12}
  ( \tau -t ) =R_{21}^{21} ( \tau - t ) \nonumber \\
  &=&
  P_{12} ( \tau - t ), \nonumber \\
  R_{21}^{12} ( \tau - t) &=& R_{12}^{21*} ( \tau - t) = [P_{12} ( \tau - t ) 
  + \cos \Omega ( \tau -t ) 
  \nonumber \\
  && + \frac{\Delta}{\Omega} \sin \Omega ( \tau - t )],
\end{eqnarray}
where
\begin{equation}
  P_{12}  ( \tau -t ) 
  = \frac{J_B^2}{\Omega^2} \sin^2 \frac{ \Omega ( \tau -t )}{2}.
\end{equation}
The time evolution of the diagonal element of 
 this matrix is
  determined by 
  the following
   differential equation
  \cite{Burshtein,Burshtein_T1}:
\begin{equation}
       \label{eqn:eqT1}
n'''( \tau ) + \frac{2}{\tau_0} n'' (\tau) 
   + ( \frac{1}{\tau_0^2} + \Delta^2 + J_B^2 ) n'(\tau)
   + \frac{J_B^2}{\tau_0} n( \tau ) = 0.
\end{equation}
Here, $n( \tau ) = \rho_{11} - \rho_{22}$, and
the initial conditions are
$ n''(0) = - J_B^2 n(0)$ and $n'(0) = 0$; and $n(0)$ 
      depends on the initial population having an
      arbitrary value
      between 1 and -1.
The time constant of the exponential decay of $n(\tau)$ is the relaxation
time of the population, $T_1$. 
When $| J_B | \ll \Delta, 1/\tau_0$ (weak coupling),
       $T_1^{-1} \sim J_B^2 \tau_0$.
When $| J_B | \gg \Delta, 1/\tau_0$ (strong coupling),
       $T_1^{-1} \sim 1/2\tau_0$.

The obtained differential equation for the off-diagonal element is     
\begin{eqnarray}
         \label{eqn:eqT2}
 && \sigma_{12}''' (\tau) + \frac{2}{\tau_0}  \sigma_{12}'' ( \tau )
  + ( \frac{1}{\tau_0^2} + \Delta^2 + J_B^2 + i \frac{\Delta}{\tau_0} )
  \sigma_{12}' (\tau)
  \nonumber \\
  &+& ( i \frac{\Delta}{\tau_0^2} + \frac{\Delta^2}{\tau_0}
  + \frac{J_B^2}{2 \tau_0} )
  \sigma_{12} ( \tau ) 
  = \frac{J_B^2}{2 \tau_0} \sigma_{21} ( \tau ),
\end{eqnarray}
where 
$ \sigma_{12} ( \tau ) = 
          \rho_{12} ( \tau ) / | \rho_{12} (0)| $ is
 the normalized off-diagonal element of the qubit density matrix.
The initial conditions are       
\begin{eqnarray}
             \label{eqn:eqT2Ini1}
 \sigma_{12}'' (0) &=& ( - \Delta^2 - \frac{J_B^2}{2} ) \sigma_{12} (0)
     + \frac{J_B^2}{2} \sigma_{21} (0), \\
              \label{eqn:eqT2Ini2}
 \sigma_{12}'(0)  &=&   - i \Delta \sigma_{12} (0), \\
              \label{eqn:eqT2Ini3}
 \sigma_{12} (0) &=& e^{ i \phi},
\end{eqnarray}
     where
     $\phi$ is the initial phase of the off-diagonal
     density matrix element.

\subsection{Analytic solutions}

Differential equation (\ref{eqn:eqT2}) with initial condition
  Eqs. (\ref{eqn:eqT2Ini1} - \ref{eqn:eqT2Ini3}) has explicit solutions:
\begin{eqnarray}
\sigma_{12} ( u ) &=& \sum_{i=1}^3 [ e^{\lambda_i u} 
+\frac{1}{\alpha} \lambda_i C_{ic}(e^{\lambda_i u}-e^{-u}) ] \nonumber \\
& & \times  ( C_{ic} \cos \phi + i C_{is} \sin \phi ) 
  + i e^{-u} \sin \phi, 
\end{eqnarray}
where $ u = \tau/\tau_0$.
For $i=1 - 3$, the coefficients are 
given by
\begin{eqnarray}
C_{ic}&=&\frac{1}{\Delta_i}((1+\beta^2)\lambda_i^2 \nonumber \\
 &+&
(1+\beta^2-2\alpha^2)\lambda_i+(1+\beta^2)\gamma^2-3\alpha^2),\\
C_{is}&=&\frac{1}{\Delta_i}\alpha(2\lambda_i^2+3\lambda_i+1+\beta^2), \\
\Delta_i&=&(\lambda_i^2+2\lambda_i+1+\gamma^2)
               (3\lambda_i^2+2\lambda_i+\gamma^2),
\end{eqnarray}
    and  the $\lambda_i's$ are the three solutions of
\begin{equation}
          \label{eqn:lambda}
   \lambda^3 + \lambda^2 + \gamma^2 \lambda + \alpha^2 = 0,
\end{equation}
where
$\alpha=\tau_0 \Delta$, $\beta=\tau_0 J_B$,
and $\gamma^2=\alpha^2+\beta^2$.

We will show one special case and three asymptotes.

$\beta=0$ {\bf isolated system}:\quad
Since 
 Eq. (\ref{eqn:lambda})
has solutions $\lambda_1=-1$ and $\lambda_{2,3}=\pm i\alpha$,
the coefficients are determined as $C_{1c}=C_{1s}=0$,
$C_{2c}=C_{3c}=\frac{1}{2}$ and
$C_{2s}=C_{3s}=\frac{1}{2i}$.
Therefore, we get 
$\sigma_{12} ( u ) = e^{ - i \alpha u + i \phi}$,
which is simply a natural rotation of the 
       off-diagonal element of
       the density matrix.

$\alpha,\beta \ll 1$ {\bf fast modulation}:\quad
In this asymptotic case, the roots of Eq. (\ref{eqn:lambda})
are
$\lambda_1\sim -1+\beta^2$ and
$\lambda_{2,3}\sim -\frac{1}{2}\beta^2\pm i\alpha$.
After determining coefficients $C_i$,
    we obtain
\begin{eqnarray}
 \sigma_{12} ( u ) \sim e^{ - i \alpha u - \frac{\beta^2}{2}u + i \phi }.
\end{eqnarray}
Therefore, the off-diagonal element of the qubit density matrix
     decays exponentially at a rate of
     $T_2^{-1} \sim \frac{1}{2} J_B^2 \tau_0$.

$\alpha,\beta \gg 1$ {\bf slow modulation}:\quad
The roots are 
$\lambda_1\sim-\frac{\alpha^2}{\gamma^2}$
and 
$\lambda_{2,3}\sim -\frac{\beta^2}{2\gamma^2}\pm \gamma i$,
so
$\Re \sigma_{12} ( u ) \sim \frac{1}{\gamma^2 }(\beta^2
e^{-\frac{\alpha^2}{\gamma^2}u}+
\alpha^2 e^{-\frac{\beta^2}{2\gamma^2}u}\cos\gamma u)\cos\phi
+\frac{\alpha}{\gamma}e^{-\frac{\beta^2}{2\gamma^2} u}\sin\gamma
u \sin\phi $ and
$\Im \sigma_{12} ( u ) \sim e^{-\frac{\beta^2}{2\gamma^2}u}
(-\frac{\alpha}{\gamma}\sin\gamma u \cos\phi+\cos\gamma u\sin\phi)$.
In particular,
for $1\ll\alpha\ll \beta$ (strong coupling limit),
\begin{eqnarray}
\sigma_{12} ( \tau ) &\sim& e^{-\frac{\alpha^2}{\beta^2}u}\cos\phi 
  \nonumber \\
 &+& i e^{-\frac{1}{2}u}\cos\beta u\sin\phi.
\end{eqnarray}
These apparent different time dependences between the real and
        imaginary parts stem from the choice of coupling
         in the form $J_B \sigma_x X/2$.
If we choose the form  $J_B \sigma_y X/2$ instead,
        the time dependences of the real and
        imaginary parts are interchanged.
In this strong coupling limit, the time evolution of $\sigma_{12} (\tau)$
      explicitly depends on its initial phase, $\phi$,
      not the simply like $e^{i \phi}$.
Therefore, as will be discussed later,
      if there are several such impurities,
      the total time evolution of $\sigma_{12} (\tau)$
       is not the simple product
      of each impurity's contribution.                
For $1\ll\beta\ll \alpha$ (weak coupling), we have
\begin{eqnarray}
\sigma ( u ) &\sim& e^{-\frac{\beta^2}{2\alpha^2}u + i \phi}.
\end{eqnarray}
In this case,
 we have exponential decay with 
 $T_2^{-1} \sim \frac{J_B^2}{2 \Delta^2} \frac{1}{\tau_0} $.

For $ \alpha, 1 \gg \beta $ {\bf weak coupling or preservative regime},
\quad
  have $\lambda_1 \sim -1 - \frac{\beta}{\alpha^2 + 1}i$,
  and 
  $\lambda_{2,3} \sim - \frac{\beta^2}{2 (\alpha^2 + 1)} 
  \pm (1+\frac{\beta^2}{2 (\alpha^2 + 1 )} ) i $,
  so $\sigma_{12} \sim e^{ - i \alpha ( 1 + \frac{\beta^2}{2 (\alpha^2 +1)})
       u- \frac{\beta^2}{2(\alpha^2+1)} u + i \phi }$.
Therefore, we again  have exponential decay with     
$T_2^{-1} \sim \frac{J_B^2 \tau_0}{2(1+\Delta^2 \tau_0^2)} $.
This coincides with the Redfield result,
      which was 
 obtained
 by perturbation theory and is justified in the
      weak coupling case, $|J_B|  \ll 1/\tau_0$
      \cite{Leggett,Slichter}.
Taking the limit $\alpha \ll 1$ further,
      we restore the result for $\alpha,\beta \ll 1$,  fast modulation.

To summarize, the dephasing rate is given by
\begin{eqnarray}
      \label{eqn:T2B}
   T_2^{-1} =
   \left \{
      \begin{array}{ll}
            J_B^2 \tau_0 / 2,
           & \quad \mbox{for $\Delta, |J_B| \ll 1/\tau_0$ } \\
             \Delta^2/J_B^2 \tau_0,
           & \quad \mbox{for the 
           real part and }\\
            & \qquad \mbox{$1/\tau_0 \ll \Delta \ll |J_B|$} \\
           \quad 1/2 \tau_0, 
           & \quad \mbox{for the imaginary par} \\
           & \quad \mbox{
            and $1/\tau_0 \ll \Delta \ll |J_B| $} \\
                        J_B^2 / 2 \Delta^2 \tau_0, 
          & \quad \mbox{for $1/\tau_0 \ll |J_B| \ll \Delta$} \\
          J_B^2 \tau_0 /2(1 + \Delta^2 \tau_0^2 )
           & \quad \mbox{For $|J_B| \ll \frac{1}{\tau_0} , \Delta$}.
       \end{array}
       \right. \nonumber \\     
\end{eqnarray}
In all regimes, $T_2 > 2 \tau_0$.

\subsection{Numerical results}
Here, we show the results of solving Eqs.
  (\ref{eqn:eqT1}) and (\ref{eqn:eqT2}) numerically.
Figure 3 shows the $\tau/\tau_0$ dependence of $n( \tau )$
    and $|\sigma_{12}( \tau )|$ 
   when $|J_B| \ll \Delta , 1/\tau_0$ (weak coupling)
   along with the asymptotic curves obtained analytically.

Figure 4 shows the $\tau/\tau_0$ dependence of $n(\tau)$ and $|\sigma_{12} 
   (\tau)|$ in the
   case of strong coupling.
It also shows the asymptotic envelope for $n(\tau)$.

Figure 5 shows the $\tau/\tau_0$ dependence of $\Re \sigma_{12} (\tau)$
    in the
   case of strong coupling.
It also shows the asymptotic curve.
As shown in Figs. 3, 4, and 5,  in the two contrasting limits,
       the numerical and analytical results
       coincide very well.       
It should be noted that we do not find Gaussian decay 
       of the off-diagonal element of the density matrix 
       for the initial regime,
       in contrast to the fluctuation in the tunneling coupling constant. 
\begin{figure}
\hspace{1truecm}
\vspace{-0.5truecm}
\center
\centerline{\epsfysize=3in
\epsfbox{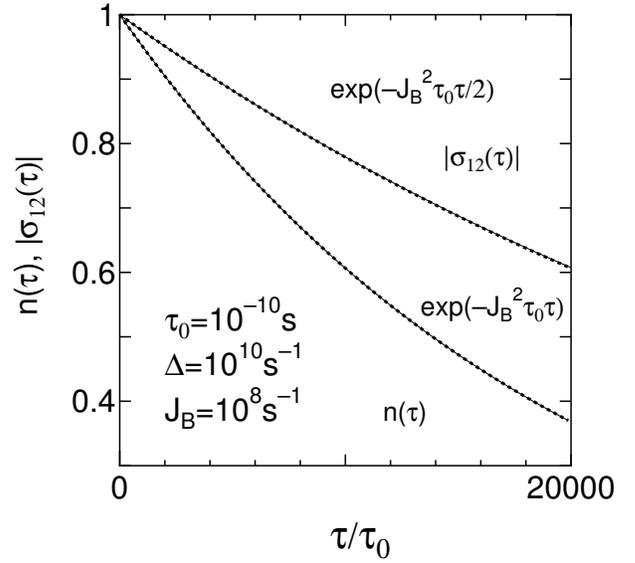}
}
\vspace{-0.5truecm}
\begin{minipage}[t]{8.1cm}
\caption{
The $\tau/\tau_0$ dependency of density matrix
 of $n(\tau)$ and $|\sigma_{12}| (\tau) $ (solid curve)
 when $J_T=0$ with  $J_B = 10^{8} s^{-1}$
 and $\Delta = 10^{10} s^{-1}$.
 Dotted lines are analytically obtained asymptotic curves,
 which is almost identical to the solid curves.
}
\label{figure1}
\end{minipage}
\end{figure}

\begin{figure}
\hspace{1truecm}
\vspace{-0.5truecm}
\center
\centerline{\epsfysize=3in
\epsfbox{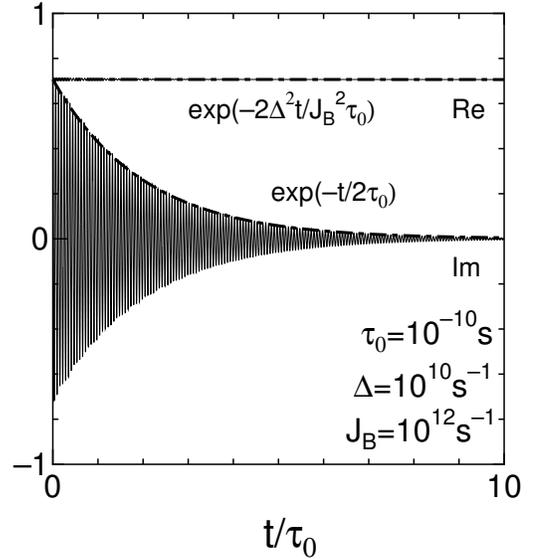}
}
\vspace{-0.5truecm}
\begin{minipage}[t]{8.1cm}
\caption{
The $\tau/\tau_0$ dependency of density matrix
 of $n(\tau)$ and $|\sigma_{12}| (\tau) $ (solid curve)
 when $J_T=0$ with   $J_B = 10^{12} s^{-1}$
 and $\Delta = 10^{10} s^{-1}$.
 Dotted line is analytically obtained asymptotic envelope curve.
}
\label{figure1}
\end{minipage}
\end{figure}
\noindent
\begin{figure}
\hspace{1truecm}
\vspace{-0.5truecm}
\center
\centerline{\epsfysize=3in
\epsfbox{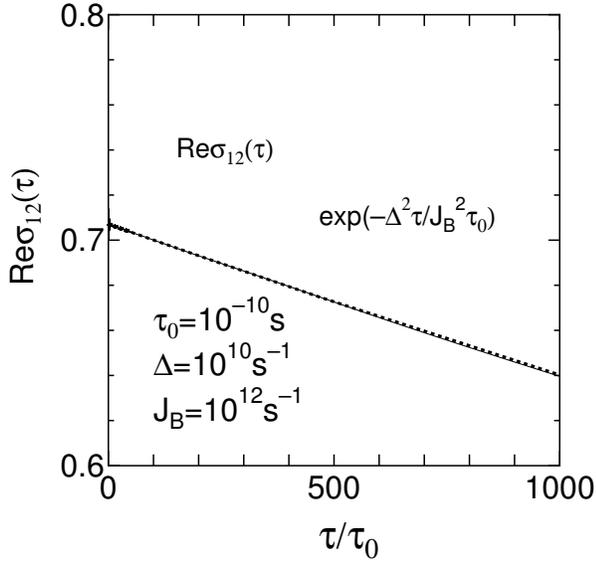}
}
\vspace{-0.5truecm}
\begin{minipage}[t]{8.1cm}
\caption{
The $\tau/\tau_0$ dependence of real part of density matrix 
       $\sigma_{12} (\tau)$ (solid line)
 when $J_T=0$ with  $J_B = 10^{12} s^{-1}$
 and $\Delta = 10^{10} s^{-1}$.
 Dotted line is analytically obtained asymptotic curve.
}
\label{figure1}
\end{minipage}
\end{figure}
\noindent
\begin{figure}
\hspace{1truecm}
\vspace{-0.5truecm}
\center
\centerline{\epsfysize=3in
\epsfbox{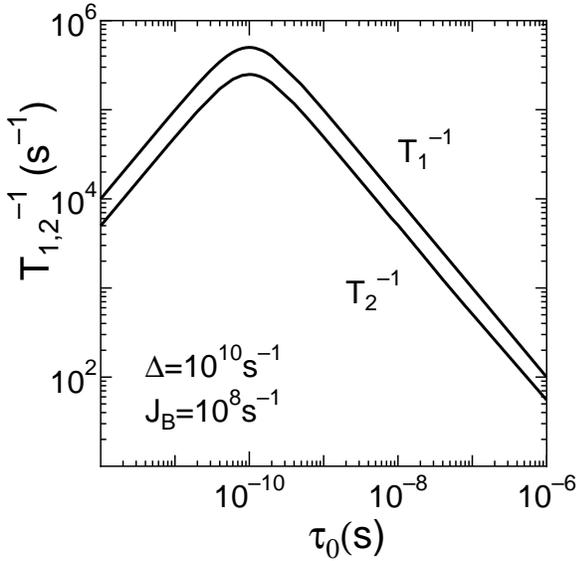}
}
\vspace{-0.5truecm}
\begin{minipage}[t]{8.1cm}
\caption{
The $\tau_0$ dependence of $T_1^{-1}$ and $T_2^{-1}$
 when $J_T=0$ with   $J_B = 10^{8} s^{-1}$
 and $\Delta = 10^{10} s^{-1}$.
}
\label{figure1}
\end{minipage}
\end{figure}

Next, we examine the $\tau_0$ dependence of the relaxation rate,
     particularly for $|J_{B}| < \Delta$.
Figure 6 shows the $\tau_0$ dependences of $T_1^{-1}$ and $T_2^{-1}$ when 
      $\Delta = 10^{10} s^{-1}$ and $J_B = 10^8 s^{-1}$.
The limits of the long and short $\tau_0$
     fit well with the analytical asymptotic given by Eq. (\ref{eqn:T2B}).
The $\tau_0$ dependence of the relaxation time is fitted as
\begin{equation}
          \label{eqn:T2BL}
T_1^{-1} = 2 T_2^{-1} = \frac{ J_B^2 \tau_0 }{ 1 + \Delta^2 \tau_0^2 }. 
\end{equation}
When $ \tau_0 < \Delta^{-1} $,
      the relaxation rates increase with $\tau_0$.
When  $ \tau_0 > \Delta^{-1} $,
      the rates decrease with an increasing $\tau_0$.
The shape of the dephasing rate as a function
   of $ \tau_0$ is explained as follows.
When $\tau_0 < \Delta^{-1}$,
     many dephasing events 
     occur during  one Rabi oscillation cycle,
     each event leads to collective disturbance.
Because the long time constant of telegraph noise leads to
       large fluctuations in the 
       variance of the rotating angle in the Bloch sphere
       during  Rabi oscillation,
       the dephasing time decreases
       with an increasing $\tau_0$.
When $ \tau_0 > \Delta^{-1} $,
    Rabi oscillation occurs over more than one cycle in time $\tau_0$
    and, in this regime, each dephasing event is
    independent.
Hence, the dephasing time increases with $\tau_0$.
The maximum around $\tau_0 \sim \Delta^{-1}$ is a kind of resonance.

\section{Discussion}

We considered the effect of electrostatic disturbance due to 
   background charge fluctuations (BCFs).
To summarize our findings, for pure dephasing ($J_T \ne 0, J_B = 0$),
         $ 2 \tau_0 < T_2 $ and $T_1 = \infty$.
For dephasing with relaxation of the 
         population ($J_T = 0, J_B \ne 0$),
         $2 \tau_0 < T_2 = 2 T_1$.

Next, we discuss the relationship between the experimentally observed $T_2$
         and our results.
In the present study,
       the dephasing time with a single background charge was
       found to be 
       longer than
       the time constant of the telegraph noise
       for both tunneling and bias fluctuations.
The observed 
       time constant of 
       a dominant random telegraph noise  
       is about 30 $\mu s$ or longer
\cite{Lyon,Fujisawa_BC},
so  a rather long dephasing time is expected. 
However, 
 in another experiment, the dephasing time was about 1 ns 
\cite{Fujisawa_PH}.        
      Therefore, a single telegraph noise source
      may  
      not be enough
      to explain the experimental results; 
      we should thus consider 
      the effect of many impurity sites or other additional effects.

We consider the effect of many impurities
      for the case in which 
      there is no correlation between background charges.
Flucuation in the tunnel coupling arises from
         the modulation of the wave function in the coupled dots.
The gradient of the electrostatic potential around
         the tunneling barrier,
         which comes from 
         an electron or hole located at an impurity site,
          leads to
         a change in the tunnel coupling
         as well as a change in the asymmetric bias
         \cite{Burkard}.
For a charge state that couples with a coupled-dot system symmetrically,
         the pure dephasing event is critical.         
For dephasing accompanied with relaxation of the population,
 the unitary operators of each impurity,
 which lead to dephasing of the qubit, are not commuting.
 However, by neglecting the higher order  $J_B \tau_0$'s in the 
 dephasing rate,
        we can take the ensemble sum of the effect of each charge state.
In such a weak coupling case,
 $(J_B \tau_0 \ll 1)$, we can use Eq. (\ref{eqn:T2BL}), 
  and  the simple  summation of the dephasing rate  
        is expressed by 
        \cite{Imry}
\begin{eqnarray} 
       \label{eqn:T2T}
       T_{2 \Sigma}^{-1} 
       &=& \int dJ_B P_{\Sigma} (J_B) \int dW P(W) T_2^{-1} ( J_B, W )
       \nonumber \\
       &=& \frac{k_B T}{ W_0} \frac{ \pi <J_B^2>_{\Sigma}}{4 \Delta}, 
\end{eqnarray}
where $W_0$ is the distribution width of the thermal activation energy
           of the charge states, 
          $P_{\Sigma} (J_B)$ is the distribution function of $J_B$,
          which depends on the relative position between
          the qubit and impurity site,
         and $<J_B^2>_{\Sigma} (=\int d J_B P_{\Sigma} (J_B) J_B^2)$
          is the sum  
          over the
          random impurities.
In the 
second equation, we assume uniform distribution
         of the activation energies of the background charges, $P(W)=1/W_0$;
         for typical cases, $ W_0 / k_B T$ is approximately 23
 \cite{Imry}. 
Use of the  perturbation method showed
           that the dephasing rate
            of a Josephson charge qubit in terms of $J_B \tau_0$
           is proportional to the inverse of $E_J$ in the limit of $E_C=0$
           \cite{Fazio,Shnirman},
           where  
           $E_J$ and $ E_C$ are the Josephson coupling constant
           and charging energy, respectively.                    
This is similar to the  estimate of $T_{2 \Sigma }^{-1}$
 in Eq. (\ref{eqn:T2T}),
 where  $E_J /\hbar $  corresponds to $\Delta$.
With a larger $\Delta$ and lower temperature,
       the dephasing rate is lower.

We next estimate the magnitude of the fluctuations, $J_B$.
The asymmetric bias fluctuation  comes from asymmetric coupling
         between the two dots and the background charge,    
which is in the form of a dipole interaction,
         $J_B \propto e^2 d \cos \theta /r^2$ for $d \ll r$,
        where $d$ is the distance between the two dots,
        $r$ is the distance between the coupled-dot system and 
        the background charge,
        and $\theta$ is the angle between them
        \cite{dipole}.
Therefore, for a smaller qubit or a charge located far from the qubit,
        the effect of bias fluctuation 
        should be less important.        
The dephasing rate is proportional to 
    $<J_B^2>_{\Sigma}$, which
    is estimated as 
\begin{eqnarray}
    \sum_i ( \frac{e^2 d}{ 4 \pi \epsilon_r \epsilon_0 \hbar r_i^2}  
    \cos \theta_i )^2
    &\sim& 
    (\frac{e^2 d}{4 \pi \epsilon_r \epsilon_0 \hbar})^2 
    N_i \int_{r_m(d)}^{\infty} r^2 dr  \nonumber \\
   & & \times \int_0^{\pi} \sin \theta
    \frac{\cos^2 \theta}{ r^4} d \theta 2 \pi \nonumber \\
    &=& (\frac{e^2d}{4 \epsilon_r \epsilon_0 \hbar})^2 \frac{4 \pi}{3}
     \frac{N_i}{r_m (d) }
\end{eqnarray}
      for the impurity sites where the dipole approximation is appropriate,
      where $r_m(d)$ is the radius
      beyond which
      the dipole approximation is valid,
      which  depends on $d$,
      $N_i$ is the density of impurity sites, and
      $\epsilon_r$ is the relative dielectric constant.            
Therefore, the total dephasing time is well defined.
The quality factor of a quantum logic gate is defined by
        the ratio of $\Delta$ to $T_{2 \Sigma }^{-1}$:
\begin{equation}
   \label{eqn:Q}
 Q= 
 \frac{\Delta}{T_{2 \Sigma }^{-1}} 
 = \frac{W_0}{k_B T} \frac{ 4 \Delta^2}{\pi < J_B^2 >_{\Sigma}},
\end{equation}
which represents how many gate operations can be done before
     the quantum coherence vanishes.
From Eq. (\ref{eqn:Q}), 
  we conclude that a large $\frac{\Delta^2}{<J_B^2>_{\Sigma}}$ 
     is needed for quantum computation. 
We estimated Q using    
      $d=0.3$ $\mu m$,
      $r_m=1$ $\mu m$, 
     $\Delta=200$ $\mu eV$ (characteristic parameter values for
     an experiment in which 
     the quantum mechanical coupling of the dots was observed in the
     frequency domain
     \cite{Fujisawa_NT} ),
     and $\epsilon_r =12.5$ (for GaAs).
To enable quantum error correction,
     the lower bound of the necessary gate quality factor was
     roughly estimated as
     $Q > 10^{6}$
     \cite{Preskill}.
Thus, 
      density of charge states  should be less than
      $5 \times 10^{6} cm^{-3} $ for fabrication.
If there is a correlation between impurities 
       (a screening effect), dephasing will be suppressed in general
\cite{Itakura_Tokura}.
It should be noted that there might be strong dephasing from
      the nearby impurities
       for which  dipole approximation is not adequate,
       even if these are only a few impurities, 
     (a few in this case).
The non-commutativity between the qubit Hamiltonian
     and environment Hamiltonian
     and the qubit backaction
      make it difficult to evaluate 
     the dephasing rate
     for strongly coupled background charge fluctuations
      in the asymmetric bias case
\cite{Fazio}.

 Finally, we discuss the  Josephson charge qubit system
\cite{Nakamura}.
Under an appropriate condition 
      (single-electron 
     charging energy $E_C$ much larger than  Josephson coupling
     energy
     $E_J$ and temperature $k_B T \ll E_J$ )
     only two charge states in the Cooper pair box (CPB)
     are important, and the 
      Hamiltonian is given by
\begin{equation}
  H = \frac{E_J}{2} \sigma_x + \frac{\delta E_C}{2} \sigma_z
   +  \frac{\hbar \sum_i J_{Ci} X_i}{2} \sigma_z,
\end{equation}
  where  
  $\delta E_C = 4 E_C ( Q_t/e - 1)$
  is the energy difference between the two charge states,
  and $Q_t$ is the total gate-induced charge in the box.
The two-charge-state basis is expressed using Pauli matrices,
   and $\hbar J_C$ is the coupling strength
  between the qubit and the background charge,
  which induces fluctuation in the charging energy. 
The  $E_J/\hbar$ corresponds to the asymmetric bias $\epsilon$,
  and $E_C/\hbar$ corresponds to $\Delta$.
Here,  
$E_C \simeq 122$ $\mu$eV, and $E_J \simeq 34$ $\mu$eV 
\cite{Nakamura_CE};
  if we can neglect  $E_J$,
  the pure dephasing event is critical.
In  pure dephasing,
         the effect of a large number of
          impurities is obtained by simply summing
           the dephasing rates,
          because  $H_{qb}$ and interaction Hamiltonian
          commute.
When the background charge and CPB interact,
    the charging energy in the CPB fluctuates.
The spectrum of  the fluctuation is given by
\begin{eqnarray}
        S_{\Delta E} ( \omega ) &=& \int d t   e^{i \omega t}
        \sum_{ij} \hbar^2 < J_{Ci} X_i (t) J_{Cj} X_j (0)> \nonumber \\
        &=& \sum_i 
        \frac{ \hbar^2 J_{Ci}^2 \tau_{0i}}{1 + \omega^2 \tau_{0i}^2}
        \nonumber \\
        &=& \int d J_C P(J_C) \int d \tau_0 P (\tau_0)
        \frac{ \hbar^2 J_{C}^2 \tau_{0}}{1 + \omega^2 \tau_{0}^2}
        \nonumber \\
        & \simeq & \frac{k_B T}{W_0} \frac{ \pi \hbar^2 <J_C^2>}{ 2 \omega},
\end{eqnarray}
where $<J_C^2>=\int d J_C P (J_C) J_C^2$,
     and we take an ensemble average over 
     the activation energy, as was done in the  coupled-dot system. 
The spectrum of the charging energy fluctuation was experimentally found to be
     $ S_{\Delta E} ( \omega ) = (\frac{4 E_C}{e} )^2 \frac{\alpha}{\omega}$,
     where $\alpha = ( 1.3 \times 10^{-3} e )^2$ \cite{Nakamura_CE}.
From  this  estimation, $< J_C^2 > \simeq 4.6 \times 10^{23} s^{-2}$
     for $20$ $mK$.     
For  an initial regime, the envelope of Rabi oscillation  shows
  Gaussian decay,
  namely, the off-diagonal element of the density matrix is given by
  $ \rho_{12} (t) \simeq \rho_{12}
   (0) {\rm exp} ( - < J_C^{2} > t^2 / 2) e^{ -i E_J t / \hbar}$.  
The rate of Gaussian decay is given by $\sqrt{<J_C^2>/2}$.
For the above value  of $<J_C^2>$, 
         the time constant of the Gaussian decay is given by 
         $3.6 \times 10^2$ $ps$,
         which
       is consistent with  the experimental finding of $150$ $ps$,  
       \cite{Nakamura_CE}.
Note that in the Gaussian regime, the time  constant does not depend on 
    the temperature.
Numerical calculation \cite{Fazio} also
       suggests this type of Gaussian decay for the pure dephasing case.

At the charge degeneracy point, namely $\delta E_C = 0$,
    dephasing with relaxation of the population occurs.
In this case,  the dephasing rate is estimated 
    using the same value of $<J_C^2>$:
\begin{equation}
    T_2^{-1} = \frac{\pi }{4 } \frac{k_B T }{W_0}
     \frac{<J_C^2>}{E_J},
\end{equation}    
so $T_2 \simeq 0.28$  $\mu s$ for $T=20$ $mK$.
In a recent experiment, a  longer  
      coherence time of $ 0.50$ $\mu s$  was found when  
      the saddle point of the ground state energy was used
      as a function of $Q_t$ and the flux
      \cite{Vion}.

\section{SUMMARY}

We examined the effect of the fluctuation of
         a single charge in an impurity site
         on a qubit.
Using the method of stochastic differential equations, 
         we calculated
         the 
         time evolution of the
         ensemble averaged density matrix of the qubit and
         obtained analytical 
         results for various conditions.         
The dephasing time, $T_2$, was always
         longer than the time constant of the random telegraph noise 
         for both tunneling 
         and bias fluctuations.
For bias fluctuation, $T_2$ was twice  the relaxation time
        of the population in the weak coupling case.
To suppress the bias fluctuation,
         the coupled dots should be positioned closer together 
         or the tunnel coupling should be made stronger.  
We also 
        investigated the case in which many impurity sites
        are
        distributed
        and examined
        the gate quality factor.
For pure dephasing, 
    which corresponds to a Josephson charge qubit
    experiment,
    the Gaussian decay of the off-diagonal element of the density matrix 
    dominated.
The present results can be applied to other quantum two-level systems
        in which there is telegraph-type fluctuations.

{\bf Acknowledgements}
We thank J. S. Tsai, T. Tanamoto, N. Imoto, M. Koashi, H. Nakano,
T. Fujisawa, Y. Hirayama, T. Hayashi, C. Uchiyama and K. Harmans 
 for their stimulating discussions.
This work was partly supported by CREST-JST, 
  Interacting Carrier Electronics Project.




\end{multicols}

\begin{references}
\bibitem{comment}
In the present study, dissipation means the relaxation of 
       the population of bonding-antibonding states.
On large time scales, the qubit state becomes  mixed with
       equal populations of bonding and anti-bonding states,
       this relaxation is not an energy relaxation.
\bibitem{dipole}
In the present study, we assume the interaction between the coupled-dot system
     and charge state to be dipole-monopole coupling.
When the electron reservoir screens the charge state effectively,
       the charge state and the reservoir act on the 
       system as a dipole.
The interaction type is dipole-dipole,
     and the interaction strength decreases more rapidly 
     as  the distance
     between the charge state and system increases.
A longer  dephasing time is thus expected compared with that of
     a dipole-monopole interaction.
\end{references}
\end{document}